# Ionospheric observations of Underground Nuclear Explosions (UNE) using GPS and the Very Large Array

Jihye Park[1], Joseph Helmboldt[2], Dorota A. Grejner-Brzezinska[1], Ralph R. B. von Frese[1], Thomas Wilson[2]

[1] The Ohio State University, Columbus, Ohio, USA

[2] US Naval Research Laboratory

Mailing address: Jihye Park, Riverside Dr. Columbus OH 43221, USA

Email: jihyepark1@gmail.com

## ABSTRACT

Observations from GPS receivers and the Very Large Array (VLA) radio telescope recorded traveling ionospheric disturbances (TID) from underground nuclear explosions (UNEs), detonated in September 1992. The slant TEC (STEC) data derived from GPS observations were processed for all ray paths to isolate TIDs. For the TIDs from the Hunters Trophy test on 18 September 1992 and the Divider test on 23 September 1992, the propagated mean velocities of the TIDs were about 573m/s and 740m/s with standard deviations of 85m/s and 135 m/s, respectively. For the VLA observations, the spectral analysis produced three-dimensional fluctuation spectral cubes for the Hunters Trophy event. The arrival time of the TID at the VLA implied a propagation speed of 570-710 m/s. This study suggests the global availability of GNSS tracking networks and new low-frequency (VHF) radio telescopes may offer a method of UNE detection and characterization, which could complement the International Monitoring System (IMS).

## 1. INTRODUCTION

Underground nuclear explosions (UNEs) affect the electron density of the ionosphere by creating acoustic-gravity waves (AGW), which sets the ionospheric plasma into motion [Hines, 1967; Park, 2012; Yang et al., 2012]. The consequence of this ionospheric status is a traveling ionospheric disturbance (TID) that propagates through the ionosphere. The ionosphere can be observed by various techniques, including ground-based ionosondes, High Frequency Radars, Doppler radar systems, dual frequency altimeters, radio telescopes, and the Global Positioning System (GPS), or, in broader terms, the Global Navigation Satellite System (GNSS). This paper focuses on the ionospheric detection of UNEs using GPS and the Very Large Array (VLA) radio telescope.

In the twenty-first century, there were two UNEs in 2006 and 2009 conducted by North Korea. Recent studies [*Park et al*, 2011; *Yang et al.*, 2012] detected the signature of the 2009 UNE by observing the ionosphere using GPS measurements. *Park et al.* [2011] detected the traveling ionospheric disturbances at multiple GNSS stations near the UNE, and determined the epicenter under the assumption of a constant propagation velocity. The statistical analysis of the array signature supports the uniqueness of the event.

A complementary technique for detecting similar and potentially weaker ionospheric disturbances using observations of cosmic sources with radio telescopes has yet to be similarly applied to the detection of UNEs. *Helmboldt et al.* [2012a,b] and *Helmboldt and Intema* [2012] have shown that interferometric radio telescopes like the VLA can be powerful tools for characterizing ionospheric fluctuations over a wide range of amplitudes and size scales. In this paper, we combine these VLA-based techniques with the GPS-based approach described by *Park et al.* [2011] to investigate the signature of any UNE in the ionosphere jointly observed by both GPS and the VLA. For this case study, we selected one of the recent UNEs in the U.S. for which contemporaneous GPS and VLA data were available. The U.S. has conducted 1,032 nuclear tests from 1945 to 1992 according to the Comprehensive Nuclear-Test-Ban Treaty (CTBTO; www.ctbto.org). Consequently, we focused on the U.S. UNEs in 1992 because during earlier epochs, GPS tended to be less stable. Table 1 summarizes two most recent U.S. UNEs in September 1992: Hunters Trophy and Divider.

We detected the signature of the Hunters Trophy test from GPS and VLA observations. The TID signature of the Divider test was only found using GPS data because no suitable VLA observations were conducted during this UNE. This study briefly presents the detection/analysis methods of UNE-induced TIDs observed by GPS and VLA respectively, and shows the results from them in reasonable agreement with one another.

## 2. GPS

### 2.1. GPS TEC and TID

GPS is a ranging system that transmits radio signals from satellites in space to an unlimited number of receivers on the Earth's surface. GPS has been available since late 1980's and currently ensures worldwide coverage with high spatial and temporal sampling rates. One of the by-products of GPS observations is the ionospheric delay in the GPS signal. The ionospheric delay can be processed into total electron content (TEC) along the slant signal path, referred to as the slant TEC (STEC). By observing and analyzing STEC, various geophysical events that disturb the ionosphere, such as earthquakes [*Zaslavski et al.*, 1998; *Liu et al.*, 2009], tsunamis [*Artru et al.*, 2005; *Galvan et al.*, 2012], explosive volcanic eruptions [*Heki*, 2006], mine blasts [*Calais et al.*, 1998], and underground nuclear explosions (UNEs) [*Park et al.*, 2011; *Yang et al.*, 2012] can be detected. In our previous study [*Park et al.*, 2011; *Park*, 2012], the STEC was computed using a dual frequency GPS signal observed at the GPS stations near the UNE. We applied the numerical derivative method to eliminate the main trend of STEC associated with the Sun's diurnal cycle and the changing geometry of the corresponding satellites, which isolates the TID signature induced by a specific source. The detailed procedure can be found at [*Park*, 2012].

In order to connect the occurrence of TIDs to a specific event, they must be distinguished from other, naturally occurring TIDs generated by phenomena such as geomagnetic storms. One of the best ways to do this is to look for the array signature of a TID that propagates from a point source. The probability of the array signature is a function of the power of the number of TID observations. For example, the probability of the experiment for 2009 North Korean UNE by *Park et al.* [2011] was approximately $1/(19\times10^9)$. Using one of the TIDs from 2009 UNE as a reference signal, we also detected the 2006 North Korean UNE [*Park et al.*, 2012], which suggests that the TID waves induced by the two UNE's had similar properties [*Park*, 2012]. In this paper, we investigated more cases of UNEs, described in Table 1, which are likely verified with another sensor, the VLA, in the following section.

**2.2. Hunters Trophy and Divider**

GPS offers a well-established, worldwide infrastructure with 439 permanent, continuously-operating GPS stations in the International GNSS Service (IGS; www.igs.org) global tracking network. The IGS tracking network data and the orbit information since 1992 are available in their archive. We collected the GPS data available in the U.S. on the days of the Hunters Trophy and Divider tests to extract STEC of all possible ray paths, which are between every GPS station and every GPS satellite, applied the numerical third order horizontal 3-point derivative method [*Park et al.*, 2011] to compute STEC derivatives. The TID-like waves were automatically detected using the correlative property of TIDs that are initially generated from the same type of source. In this study, we used the TID induced by the 2009 North Korean UNE in our previous study [*Park et al.*, 2011] as a reference signal. Figures 1 and 2 show the relationship between the travel time and distance of the detected TID signal that fit to the straight-line model. This implies that the TID waves induced by these point sources in both cases propagated with certain constant speeds.

Figure 1 shows that the data points are reasonably well represented by the travel time-distance curve, which is the first order polynomial regression model. The travel distance used here is the slant distance from a UNE to the ionosphere's pierce point (IPP), where the ray connecting the receiver to the GNSS satellite intersects the peak of ionosphere at 300 km altitude. Within the 95% confidence interval, eight out of ten data points were included. The approximate propagation velocity of the TID from the Hunters Trophy UNE was about 573.00 m/s with the standard deviation of 84.68 m/s. The VLA observations described in section 3 also detected the effects of the Hunters Trophy TID at an arrival time that is consistent with this GPS detection result. Figure 2 shows the travel time-distance curve of the Divider test. The data points in Figure 2 are also well represented by the straight line fit with a TID propagation velocity of 739.76 m/s with the standard deviation of 134.50 m/s. The TID propagation velocities from these polynomial regression models of the two events in Figures 1 and 2 were within a range of a medium scale TID (MSTID), between several hundreds meters per second and less than a kilometer per second [*Francis*, 1974]. It should be emphasized that the propagation velocities in this study are relatively faster than another recent study of the 2009 North Korean UNE [*Park et al.*, 2011], which was about 273 m/s.

It is also interesting that the explosive yield of the 2009 North Korean UNE was 2.2-4.8 kT, an order of magnitude lower than the yields for the Hunters Trophy and Divider tests. It is possible that because the period/wavelength of the AGWs generated by these events is correlated with yield [e.g., *Herrin et al.*, 2008], the AGWs from the 1992 US tests were able to penetrate higher into the ionosphere because they would be less severely affected by viscous dissipation, which is roughly proportional to the square of the wavenumber [i.e, $\sim k^2$; *Hines,* 1960]. At F-region heights (~300-400 km), the temperature is high enough that the sound speed is comparable to the speeds detected for the Hunters Trophy and Divider events (within $\pm 1\sigma$). The slower speed of the 2009 North Korean UNE is more consistent with the typical sound speed for the E-region (~100 km). Additional studies would be required to determine if there is a consistent relationship between UNE yield and the speed of the corresponding TID.

## 3. VLA

### 3.1. The VLA and the Ionosphere

The Very Large Array (VLA) is a radio-frequency interferometer located in New Mexico (34°04'43.5"N, 107°37'05.8"W), consisting of 27 dish antennas, each 25-m in diameter and 27, 25-m dish antennas configured in an inverted "Y" with "arms" extending to the north, southeast, and southwest. The antennas are periodically cycled through four configurations, A, B, C, and D, spanning 36, 11, 3.4, and 1 km, respectively. The VLA has observing bands between 1 and 50 GHz and prior to 2008, had a separate VHF system with two bands centered at 74 and 330 MHz. A new wider-band VHF system is currently being commissioned. The VHF bands and L-band (1.4 GHz) are significantly affected by the ionosphere.

Operated as an interferometer, the VLA measures the correlation of complex voltages from each unique pair of antennas, or "baseline," producing what are referred to as visibilities. While each antenna is pointed at the same cosmic source, each antenna's line of sight passes through a different part of the ionosphere, and thus the measured visibilities have an extra phase term added as a consequence of the difference in ionospheric delays. This extra phase term is therefore proportional to the difference in the STEC along the lines of sight of the two telescopes. This makes the interferometer sensitive to the STEC gradient rather than STEC itself, which renders it capable of sensing both temporal and spatial fluctuations in STEC.

The antenna-based phases errors introduced by both the ionosphere and the VLA instrumentation are simultaneously solved for using a technique referred to as self-calibration [*Cornwell and Fomalont*, 1999]. Self-calibration uses a model of the distribution of intensity on the sky for a particular cosmic source (a simple point-source model is often sufficient) to remove from the visibilities any phase variations due to the source itself. Then, the procedure uses a standard non-linear least squares fit (typically employing a gradient search) to solve for each antenna-based phase error using all baselines. Since for $N$ antennas, there are $N(N-1)/2$ unique baselines, this is typically a very over-determined problem (e.g., for the VLA, there are 27 antennas and 351 baselines).

In a series of papers, *Helmboldt et al.* [2012a,b] and *Helmboldt and Intema* [2012] demonstrated how one could use self-calibration-determined phase corrections to characterize ionospheric fluctuations. Briefly, the instrumental contributions to the antenna phase errors are known to be quite stable in time and can be removed from the data by simply de-trending it using a sliding boxcar filter, a linear fit, or some combination of the two. What is then left for each antenna is a times series of ΔTEC, the difference in STEC between its line of sight and that of a reference antenna whose phase is arbitrary and set to zero. *Helmboldt et al.* [2012a] demonstrated that this could be done to a precision of <0.001 TECU with the VLA when observing a bright source in either of the VHF bands.

Because of the Y-shape of the array, these ΔTEC measurements cannot be used to numerically solve for the two-dimensional STEC gradient over the array. However, *Helmboldt* [2012a] showed that most of the structure in the STEC surface is recovered with a second order, two-dimensional polynomial fit. Using such a fit, one can construct a time series of the full STEC gradient over the array and a three-dimensional Fourier analysis (one temporal, two spatial) can be performed. Such a spectral analysis yields power spectrum cubes, which can be examined for significant peaks to identify wave-like features and measure their wavelengths, orientations, and speeds.

### 3.2. Hunters Trophy Observations

During the time of the Hunters Trophy UNE test, the VLA was observing a series of relatively bright cosmic sources at 1.4 GHz in dual polarization mode that are ideal for self-calibration. The VLA was in its D-configuration, spanning about 1 km. The observing run (VLA program AL252) was divided into two segments, one spanning 06:30-08:20 UT and the other 16:30-18:20 UT. The latter segment contained the time of the Hunters Trophy event, which occurred 850 km to the northwest of the VLA. Each source was observed multiple times, each time for intervals of 2-15 minutes.

The data for these observations were retrieved from the VLA archives and self-calibration was performed using each source to obtain ΔTEC time series for each antenna. Within each 2-15 minute observing chunk, the ΔTEC time series were de-trended with a linear fit. This was done separately for the receivers of the right-hand and left-hand circular polarizations and by comparing the two to one another, we estimate the 1-σ precision of these measurements to be about 0.003 TECU. This is higher than that achieved by *Helmboldt et al.* [2012a] because they used a brighter source and observed at a lower frequency, 74 MHz, for which the effect of the ionosphere is nearly 20 times larger than at L-band.

The spectral analysis described above and in more detail by *Helmboldt et al.* [2012b] and *Helmboldt and Intema* [2012] was performed on each observing chunk, producing three-dimensional fluctuation spectral cubes. Specifically, the combined spectral power from both components of the STEC gradient (i.e., east-west, or "x," and north-south, or "y") was computed according to

$$P(\nu,\xi_x,\xi_y) = |FT\left\{\frac{\partial}{\partial x}STEC(t,-x,-y)\right\}|^2 + |FT\left\{\frac{\partial}{\partial y}STEC(t,-x,-y)\right\}|^2$$

where FT denotes a three dimensional Fourier transform over time, and the two horizontal spatial axes (x and y), and ν and ξ refer to temporal and spatial frequencies, respectively. The STEC partial derivatives were computed from the VLA data using second-order, two-dimensional polynomial fits (see above). The Fourier transforms are normalized such that for a plane-wave with a STEC amplitude of A and spatial frequency ξ, the combined spectral power according to this formulation is simply 2πAξ.

For those observations near and following the Hunters Trophy event, the peak spectral power and peak temporal frequency are displayed as images in Figures 3 and 4 as functions of east-west and north-south spatial frequency. In these spectral maps, up is north, right is east, and the spatial frequency increase (i.e., the wavelength decreases) with radius. In both figures, the mean power over all temporal frequencies is also shown as contours and the name of the observed source is printed at the bottom within each panel. In addition, the expected range in azimuth for waves originating from the Hunters Trophy site for altitudes between 100 and 900 km are shown on each spectral map with black lines.

From these maps, between roughly 20 and 25 minutes after the UNE, one can see that the fluctuation power dramatically increases, accompanied by the signature of wavelike structures moving nearly perpendicular to the direction from Hunters Trophy (i.e., toward the northeast and southwest). Assuming that this spike in fluctuation power is associated with the arrival of disturbances associated with the Hunters Trophy event, this implies a propagation speed of 570-710 m/s, consistent with the GPS results detailed above. The fluctuation amplitudes peak 43 minutes after the event, and then fade relatively quickly. At first, the northeastward-directed waves appear stronger, but then fade more quickly and the southwestward-directed waves begin to dominate at about 18:00 UT. The waves are relatively small, around 2 km in wavelength (or, spatial frequency of 0.5 $km^{-1}$). This is not surprising given that the VLA was in its most compact configuration and was therefore more sensitive to smaller-scale fluctuations. In addition, the necessary detrending of the ΔTEC time series over the relatively short observing period for each source (2-15 minutes each; see above) dampens any relatively large, long-period oscillations. Figure 4 shows that the periods for the detected small-scale waves are between about 4 and 15 minutes (or, temporal frequencies of about 4-15 $hr^{-1}$). With the observed wavelengths, these imply relatively slow speeds of about 2-8 m $s^{-1}$ (or 8-30 km $hr^{-1}$).

Given their characteristics, it seems likely that these waves are associated with small-scale distortions propagating along the larger wavefront generated by the Hunters Trophy explosion, to which the compact D-configuration VLA is relatively insensitive. In other words, since the D-configuration VLA is about 1-km across, the determined propagation speed implies the main wavefront passed over the VLA in <2 s. The visibilities for the VLA observations were computed using a 40-s integration period, making it difficult if not impossible to sense the main Hunters Trophy wavefront using the VLA data. Instead, the VLA observations were able to capture the effects of this wavefront on the fine-scale dynamics of the ionosphere.

## 4. SUMMARY AND DISCUSSION

It has been revealed that UNEs disturb the ionosphere, which results in TIDs, according to several studies in the literature. This paper demonstrated the TID detection induced by the US UNEs in 1992 using GPS and the VLA separately. It should be noted that VLA observation was not available during the time of the Divider UNE test, hence, only the Hunters Trophy was jointly detected by GPS and the VLA.

Section 2 described the TID detection using GPS TEC. GPS data on the days of the Hunters Trophy and the Divider tests were collected from the IGS archive. The array signature, which was formed by multiple TID observations at different GPS stations, was one of the strongest indications that the TIDs were propagated from a point source. For the Hunters Trophy and the Divider UNE tests, the array signature of TIDs at the vicinity of GPS stations was observed for each event. By applying the first order polynomial model to compute the approximate velocity of TID propagation for each UNE, the data points in Figures 1 and 2, which were the TID observations, were sufficiently fit to the model within the 95% confidence interval. The velocities of the models were 570 m/s and 740 m/s for the Hunters Trophy and the Divider respectively, and these numbers were reasonable for the propagation velocity of the MSTID. The velocity of TID due to the Hunters Trophy test was confirmed by the VLA observations described in Section 3.

Section 3 demonstrated the mechanism of ionospheric detection of the VLA radio-frequency interferometer and its ionospheric observation during the time of the Hunters Trophy test. With the self-calibration technique, ionospheric fluctuations resulting from the Hunters Trophy event were detected. The full TEC gradient over the antenna array was computed by applying a second order, two-dimensional polynomial fit. In addition, a temporal and spatial fluctuation spectrum of the TEC gradient was also generated with a three-dimensional Fourier analysis. Roughly 20-25 minutes after the Hunters Trophy test, the signature of waves moving nearly perpendicular to the direction from the UNE were detected, which implies a propagation speed of 570-710 m/s. These wavelike structures were likely small-scale (~2km in wavelength) distortions associated with the larger disturbance(s) generated by the UNE.

As we discussed in Section 2, the array signature of TID was found using GPS observations and this detection was supported by strong statistics. In addition, the trace of one of the UNEs in the ionosphere was detected by totally independent sensor with different detection/analyzing algorithm, the VLA, which yielded a similar propagation velocity of the TID from the UNE. This study suggests that the global availability of GNSS tracking networks combined with further algorithmic improvement of this method and new low-frequency (VHF) radio telescopes may offer a future UNE detection method, which could complement the International Monitoring System (IMS).

**ACKNOWLEDMENTS**

Basic research in astronomy at the Naval Research Laboratory is supported by 6.1 base funding. The National Radio Astronomy Observatory is a facility of the National Science Foundation operated under cooperative agreement by Associated Universities, Inc.

**Tables:**

Table 1: Event description of 1992 U.S. UNEs

| Country | USA (South Nevada) | USA (Yucca Flat) |
|---|---|---|
| Name | Hunters Trophy | Divider |
| Source Type | Underground | Underground |
| Data Time [UTC] | 1992/09/18 17:00:0.080 | 1992/09/23 15:04:0.0 |
| Latitude/Longitude [deg] | 37.207/-116.211 | 37.021/-115.989 |
| Depth [km] | 0.385 | 0.34 |
| Yield [kT] | <= 20 | <= 20 |
| Magnitude [mb] | 4.3 | 4.3 |

## Figures

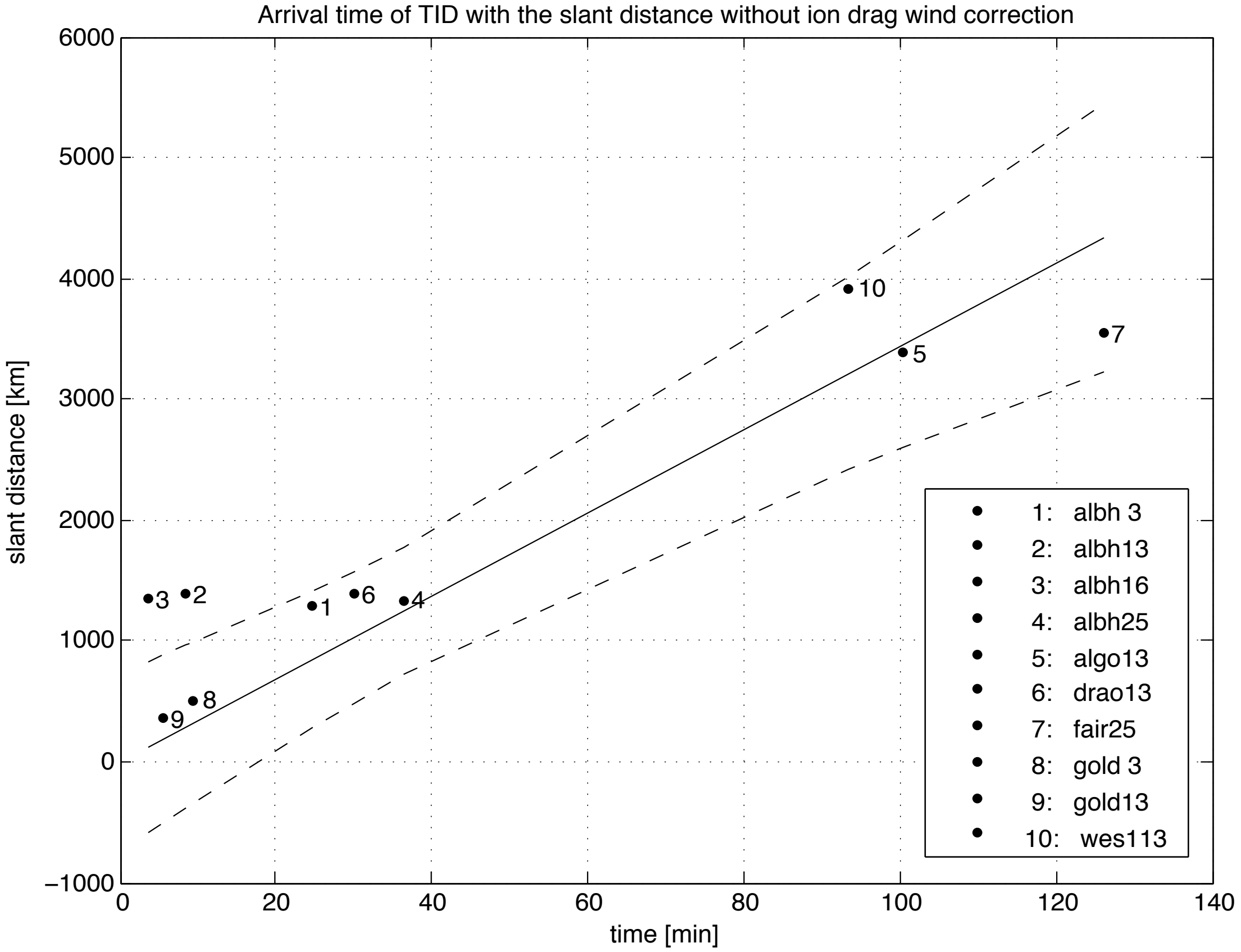

Figure 1: Travel-time data for the putative UNE (Hunters Trophy)-generated TIDs without Ionospheric wind corrections. The correlation coefficient of the data points on the fitting curve is 0.93. The propagation velocity computed from the fitting curve is 573.00 m/s with the standard deviation of 84.68 m/s.

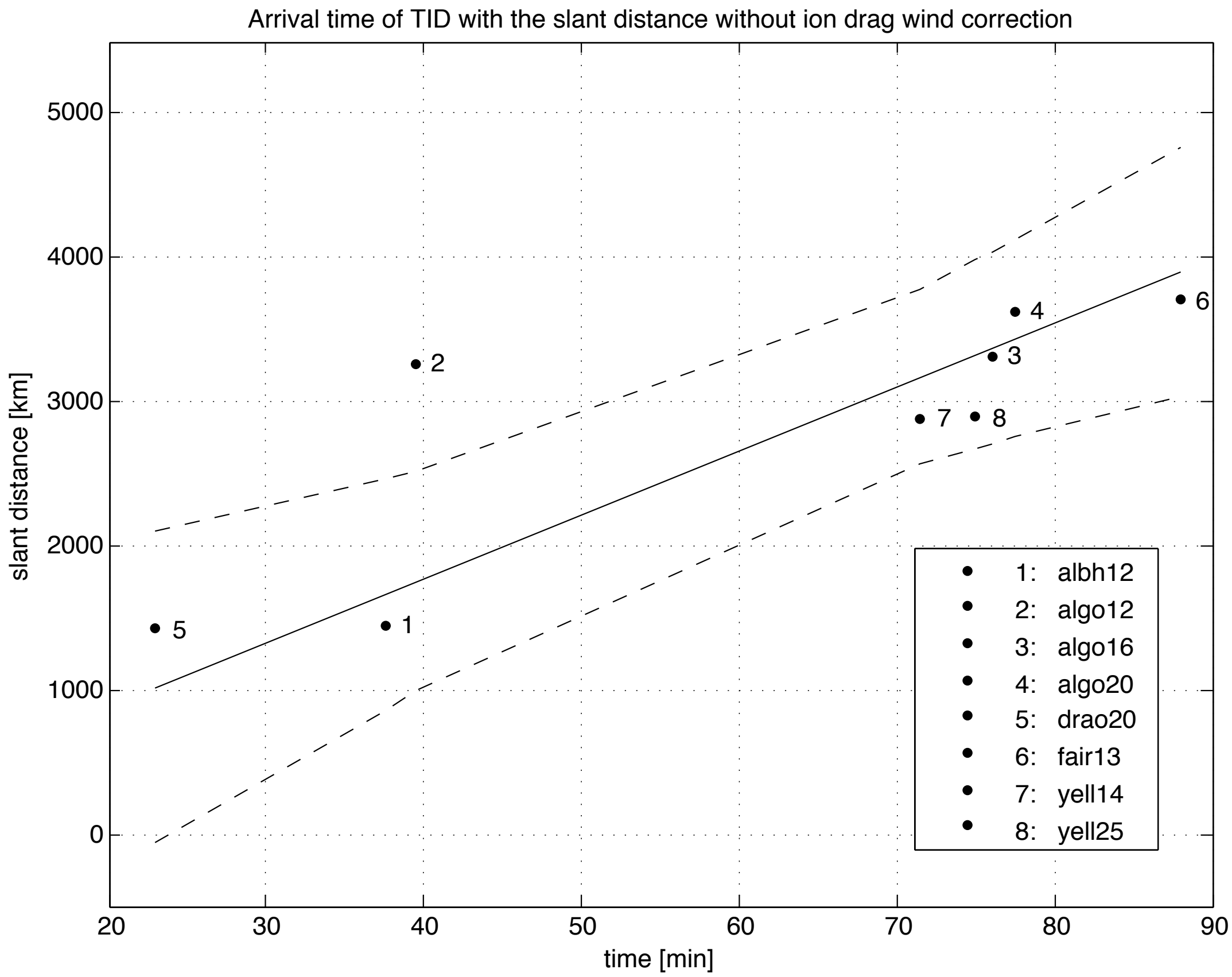

Figure 2: Travel time data for the putative UNE (Divider)-generated TIDs without Ionospheric wind corrections. The correlation coefficient of the data points on the fitting curve is 0.81. The propagation velocity computed from the fitting curve is 739.76 m/s with the standard deviation of 134.50 m/s.

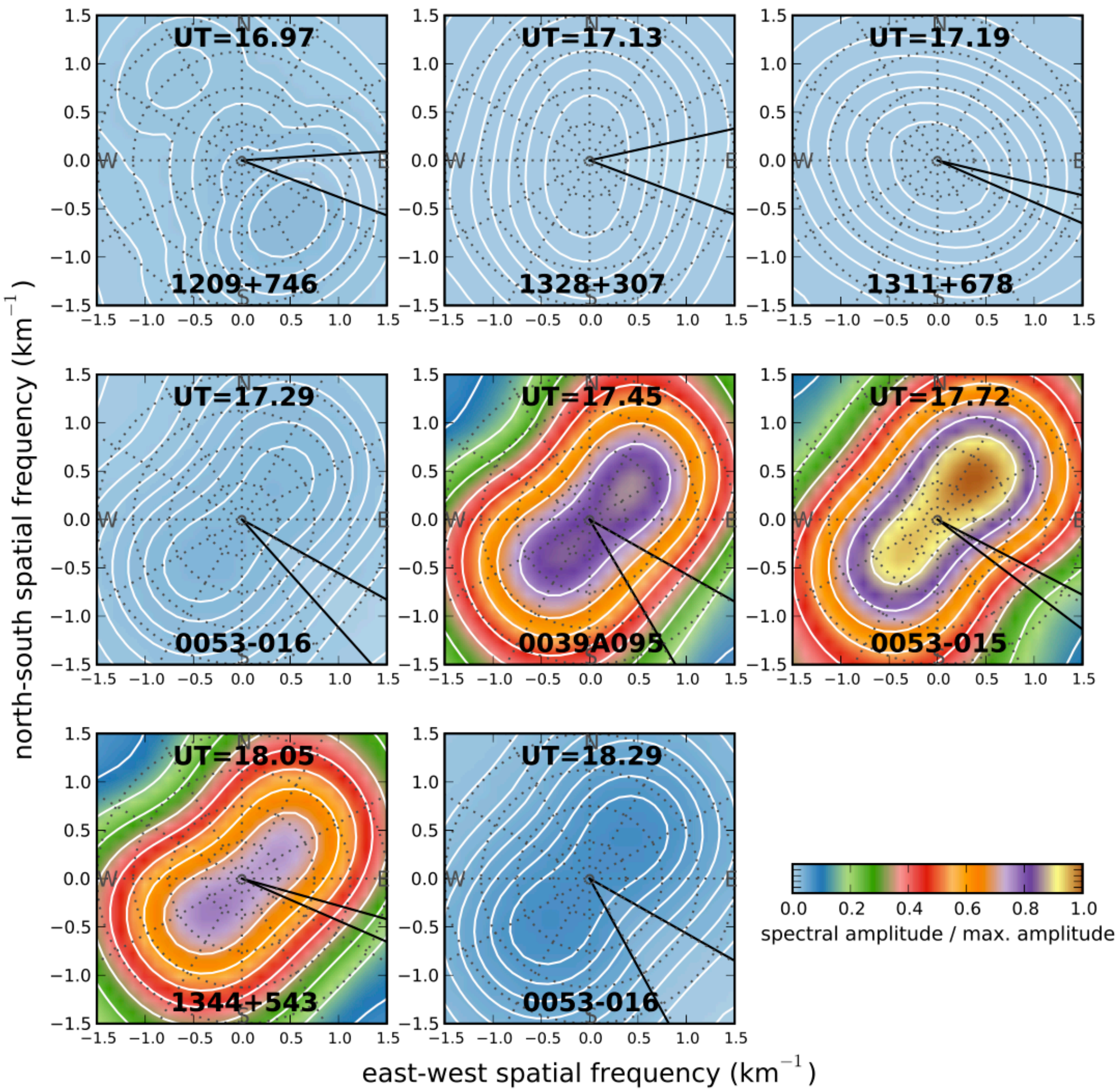

Figure 3: For each source observed near and after the Hunters Trophy event, maps of TEC gradient fluctuation power as a function of east-west and north-south spatial frequency. Here, the peak spectral power over all temporal frequencies is shown as an image in each panel with contours representing the mean power over all temporal frequencies. The range in azimuths expected for waves propagating from Hunters Trophy for altitudes between 100 and 900 km is shown in each panel with black lines.

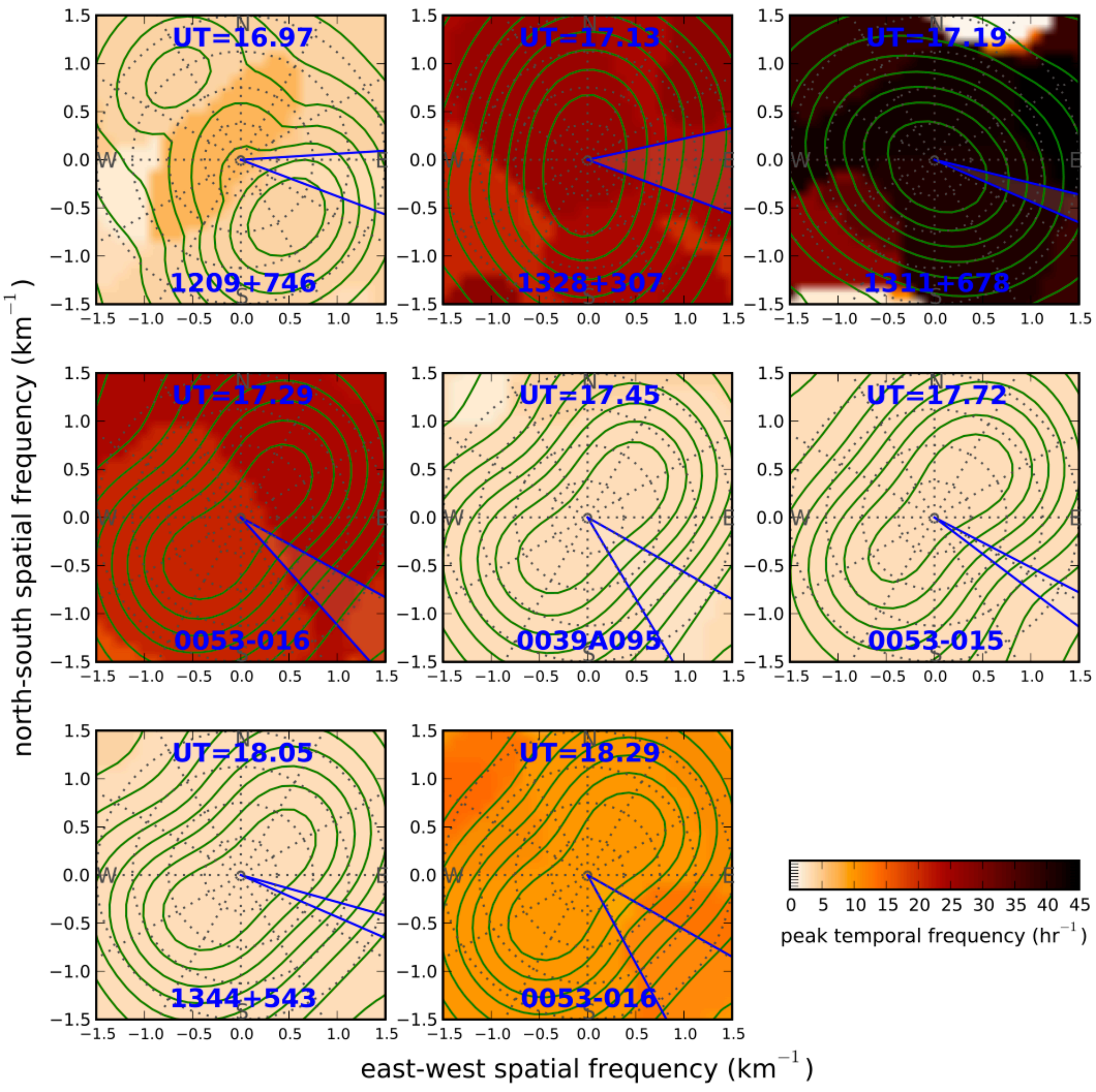

Figure 4: The same as Figure 3, but with the peak temporal frequency displayed as images.